\documentclass[%
preprint,
superscriptaddress,
 amsmath,amssymb,mathtools,
 aps,
]{revtex4-1}

\usepackage{graphicx}
\usepackage{dcolumn}
\usepackage{bm}
\usepackage{lineno,hyperref}%
\usepackage{graphicx, epsfig, amsmath, amssymb, afterpage,makeidx, setspace, subfigure}
\usepackage{color}
\usepackage[english]{babel}
\usepackage[dvipsnames]{xcolor}
\usepackage{float}
\usepackage{siunitx}

\begin{document}
\preprint{APS/PhysRevLett}

\title{Proton Bunch Self-Modulation in Plasma with Density Gradient}

\author{F.~Braunm\"{u}ller}
\affiliation{Max Planck Institute for Physics, Munich, Germany}
\author{T.~Nechaeva}
\affiliation{Belarusian State University, 220030 Minsk, Belarus}
\author{E.~Adli}
\affiliation{University of Oslo, Oslo, Norway}
\author{R.~Agnello}
\affiliation{Ecole Polytechnique Federale de Lausanne (EPFL), Swiss Plasma Center (SPC), Lausanne, Switzerland}
\author{M.~Aladi}
\affiliation{Wigner Research Center for Physics, Budapest, Hungary}
\author{Y.~Andrebe}
\affiliation{Ecole Polytechnique Federale de Lausanne (EPFL), Swiss Plasma Center (SPC), Lausanne, Switzerland}
\author{O.~Apsimon}
\affiliation{Cockcroft Institute, Daresbury, UK}
\affiliation{Lancaster University, Lancaster, UK}
\author{R.~Apsimon}
\affiliation{Cockcroft Institute, Daresbury, UK}
\affiliation{Lancaster University, Lancaster, UK}
\author{A.-M.~Bachmann}
\affiliation{Max Planck Institute for Physics, Munich, Germany}
\affiliation{CERN, Geneva, Switzerland}
\affiliation{Technical University Munich, Munich, Germany}
\author{M.A.~Baistrukov}
\affiliation{Budker Institute of Nuclear Physics SB RAS, Novosibirsk, Russia}
\affiliation{Novosibirsk State University, Novosibirsk, Russia}
% ...
\collaboration{The AWAKE Collaboration}
\noaffiliation

\sloppy 

\date{\today}
\begin{abstract}

We study experimentally the effect of linear plasma density gradients on the self-modulation of a 400\,GeV proton bunch. %
Results show that a positive/negative gradient in/decreases the number of micro-bunches and the relative charge per micro-bunch observed after 10\,m of plasma. %
The measured modulation frequency also in/decreases. 
With the largest positive gradient we observe two frequencies in the modulation power spectrum. %
Results are consistent with changes in wakefields' phase velocity due to plasma density gradient adding to the  slow wakefields' phase velocity during self-modulation growth predicted by linear theory. %
\end{abstract}

\maketitle
Self-modulation (SM) of long, relativistic charged particle bunches was proposed in order to take advantage of bunches that carry large amounts of energy (kilojoules) to drive large amplitude wakefields ($>$1\,GV/m) in long plasma~\cite{KumarFirstSMI10}. %
In the SM process the transverse component of noise wakefields from the plasma or bunch, or a transverse component externally imposed as seed, grows and transforms the long bunch into a train of micro-bunches through periodic focusing and defocusing. %
In a neutral plasma with electron density $n_{e0}$ the periodicity of the micro-bunches is given by the plasma electron angular frequency: $\Delta$$t$=2$\pi/\omega_{pe}$, where $\omega_{pe}$$=$$\left(\frac{n_{e0}e^2}{\epsilon_0 m_e}\right)^\frac{1}{2}$~\cite{bib:cst}. %
Micro-bunches are shorter than $\Delta t$. %
The incoming bunch is considered as long when its duration is $\gg$$\Delta t$. %
The SM of a 400\,GeV, 230\,ps rms duration proton bunch in a 10\,m-long plasma with density in the 10$^{14}$-10$^{15}$\,cm$^{-3}$ range (11$\ge$$\Delta$t$\ge$3.5\,ps) was recently demonstrated experimentally~\cite{bib:karl,bib:turner}. 
Results show that the modulation frequency observed after the plasma is equal to the plasma electron frequency inferred from the singly-ionized rubidium (Rb) vapor density~\cite{bib:karl}, and that the SM process grows both along the bunch and plasma~\cite{bib:turner}, all as expected from theory. %
Transition between the instability regime and the seeded regime was also observed~\cite{bib:fabianSSMSMI}, demonstrating the ability to control the SM process, a necessary condition for practical accelerators. %

It was recognized in two simultaneous theory/simulation papers~\cite{bib:phasevelocity,bib:schroeder} that during growth of the SM process the wakefields' phase velocity is slower than the speed of the 400\,GeV proton bunch. %
The wakefields' phase velocity reaches essentially the velocity of the bunch only after saturation of the SM process. %
Both papers focused on the effect of slow wakefields' phase velocity on electron acceleration during wakefields' growth. %
Both also refer to a possible solution that was devised for the case of a long laser pulse self-modulating in a plasma: a plasma density gradient to manipulate the effective wakefields' phase velocity~\cite{bib:pukhov2}. %
Plasma density gradients were proposed to overcome dephasing limitations on energy gain in laser wakefield accelerators~\cite{bib:kats}. %
One can expect a plasma density gradient to also affect the SM process of a particle bunch, 
perhaps leading to larger amplitude wakefields after saturation. %
 
In this {\emph Letter} we focus on the effect of plasma density gradients on the bunch that drives wakefields. %
In the experiment, we measure the effect of linear density gradients on the parameters of a 400\,GeV proton bunch after 10\,m of propagation in plasma. %
We show that the time-resolved train of micro-bunches resulting from the SM process becomes longer/shorter and that the relative charge per micro-bunch exiting the plasma in/decreases as the gradient is made more positive/negative, in practice with density larger/smaller at the plasma exit %
than at its entrance. %
An independent time integrated charge diagnostic confirms the increase in charge remaining near the bunch axis with increasing density gradient value $g$. %
We show with two diagnostics that the proton bunch modulation frequency, equal to the plasma frequency without gradient~\cite{bib:karl}, in/decreases with in/decreasing gradient values. %
At large gradient values ($\left|g\right|$$>$1\%/m) we observe saturation and for $g$=+1.9\%/m two frequencies appearing in the spectrum of the modulated bunch. %

Changes in charge and modulation frequency are consistent with 
change in phase velocity of the wakefields. 
The larger/smaller charge measured with positive/negative gradient is consistent with the wakefields' phase velocity being slower than that of protons, or negative in the protons' reference frame. %
Results presented here have been obtained in the AWAKE experiment~\cite{MuggliPPCF17}. %

\begin{figure*}[!ht]
\centerline{\includegraphics[width=0.8\linewidth]{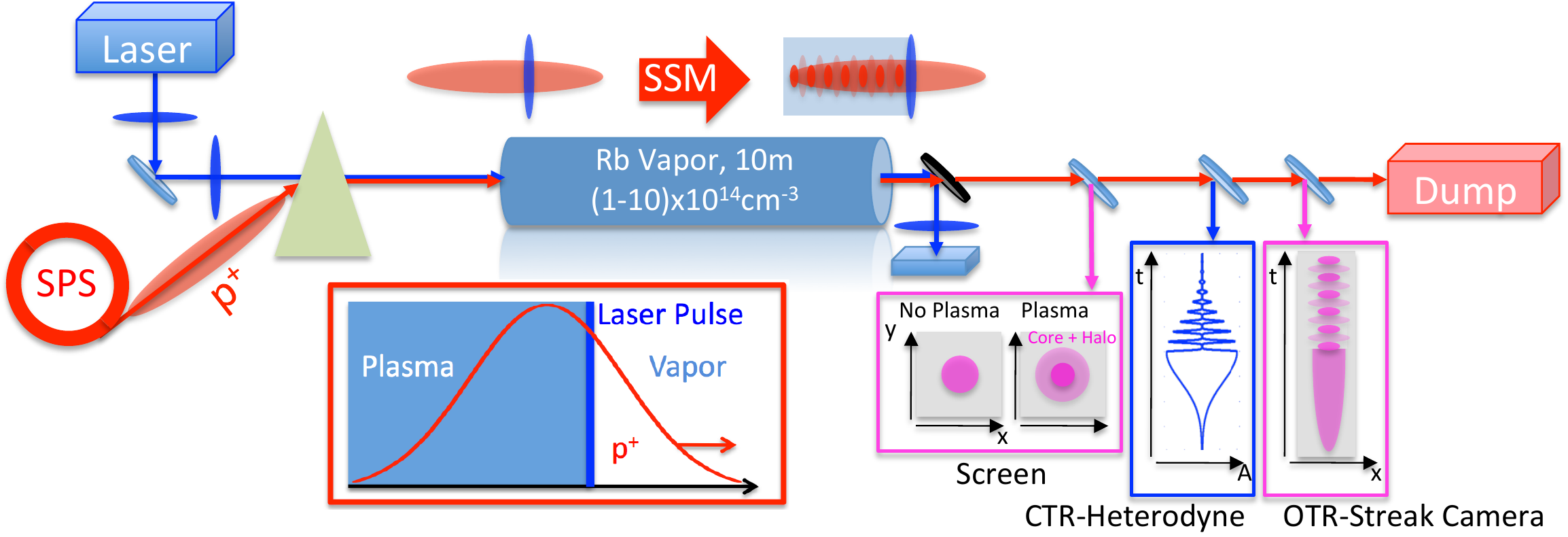}}
 \caption{\label{figSetupAWAKE}Schematic of the AWAKE experiment (not to scale).
 Red frame inset: schematic of the ionizing laser pulse and proton bunch timing along the plasma. %
}
\end{figure*}

A schematic of the AWAKE experimental setup is shown in Fig.~\ref{figSetupAWAKE}. %
The 400\,GeV proton bunch from the CERN Super Proton Synchrotron with a population of (2.98$\pm$0.16)$\cdot$10$^{11}$ particles is focused to a \SI{200}{\micro\m} rms transverse size near the entrance of the plasma. %
It has a rms duration $\sigma_{t}$=230\,ps. %
A 120\,fs-long laser pulse with an energy of $\sim$110\,mJ creates the plasma by singly ionizing a Rb vapor to form a cylinder of plasma, $\sim$1\,mm in radius, over the 10\,m-long vapor column. %
The sharp ($\ll$2$\pi/\omega_{pe}$$\cong$8.3\,ps for the plasma density used here), relativistic ionization front created by the laser pulse provides seeding for the SM process~\cite{bib:karl,bib:fabianSSMSMI}. %
The SM process and its phase are reproducible from event to event, i.e., in the seeded self-modulation (SSM) regime~\cite{MuggliPPCF17,bib:fabianSSMSMI}. %
For this purpose the laser pulse is co-propagating within the proton bunch, at a time $t_{seed}$=128\,ps (0.56$\sigma_{t}$) ahead of the bunch center (see Fig.~\ref{figSetupAWAKE} Inset). 
Experiments were performed with a fixed vapor density of $n_{Rb}$=(1.804$\pm$0.004)$\cdot10^{14}\,$cm$^{-3}$ at the plasma entrance, varying density at its exit. %
The vapor density gradient is controlled by the temperature of two reservoirs evaporating Rb into the source~\cite{bib:oz14, GennadyVapourSource18}. %
We precisely measure the vapor density ($<$0.5\%) 
at both ends with white-light interferometry~\cite{ErdemRbDiagNIMA16}.
For small density differences between the two ends ($\left|g\right|$$\le$1.9\%/m over 10\,m at densities $\le$10$^{15}$\,cm$^{-3}$), vapor flow is small enough so that the density gradient can be considered as linear along the source ($z$)~\cite{GennadyVapourSource18}. %
The density gradient ($n_{e}(z)$=$n_{e0}\left(1+\frac{g}{100}z\right)$) is therefore calculated by dividing the difference in density measured at both ends by the length of the vapor column (10\,m). %
We demonstrated in Ref.~\cite{bib:karl} that laser ionization yields a plasma density equal to the Rb vapor density within 0.1\% 
Therefore, though we measure Rb vapor densities, hereafter we quote the corresponding plasma density and gradient values ($n_{e0}$=$n_{Rb})$. %

Three diagnostics provide information about the self-modulated proton bunch after the plasma (see Fig.~\ref{figSetupAWAKE}). %
A scintillating screen located $\sim$2\,m downstream from the plasma exit yields the time-integrated transverse population density distribution of protons~\cite{MarleneTwoScreen17}. %
One meter downstream, the bunch train enters a metallic screen where it produces coherent transition radiation (CTR). %
We send the CTR to a heterodyne frequency measurement system that yields modulation frequency along the bunch (spectrogram~\cite{CTRdiagEAACNiMa18}). %
Protons then produce backward (incoherent) optical transition radiation (OTR) when entering another nearby metallic screen~\cite{bib:karlrsi,bib:karl}. %
We image the OTR onto the entrance slit of a streak camera that yields a time-resolved image of the proton bunch transverse density distribution in a $\sim\SI{74}{\micro\m}$-wide slice about the bunch propagation axis (bunch rms width $\sigma_{r0}$$\cong$$\SI{540}{\micro\m}$ without plasma). %
\begin{figure}[!ht]
\centerline{\includegraphics[width=0.8\linewidth]{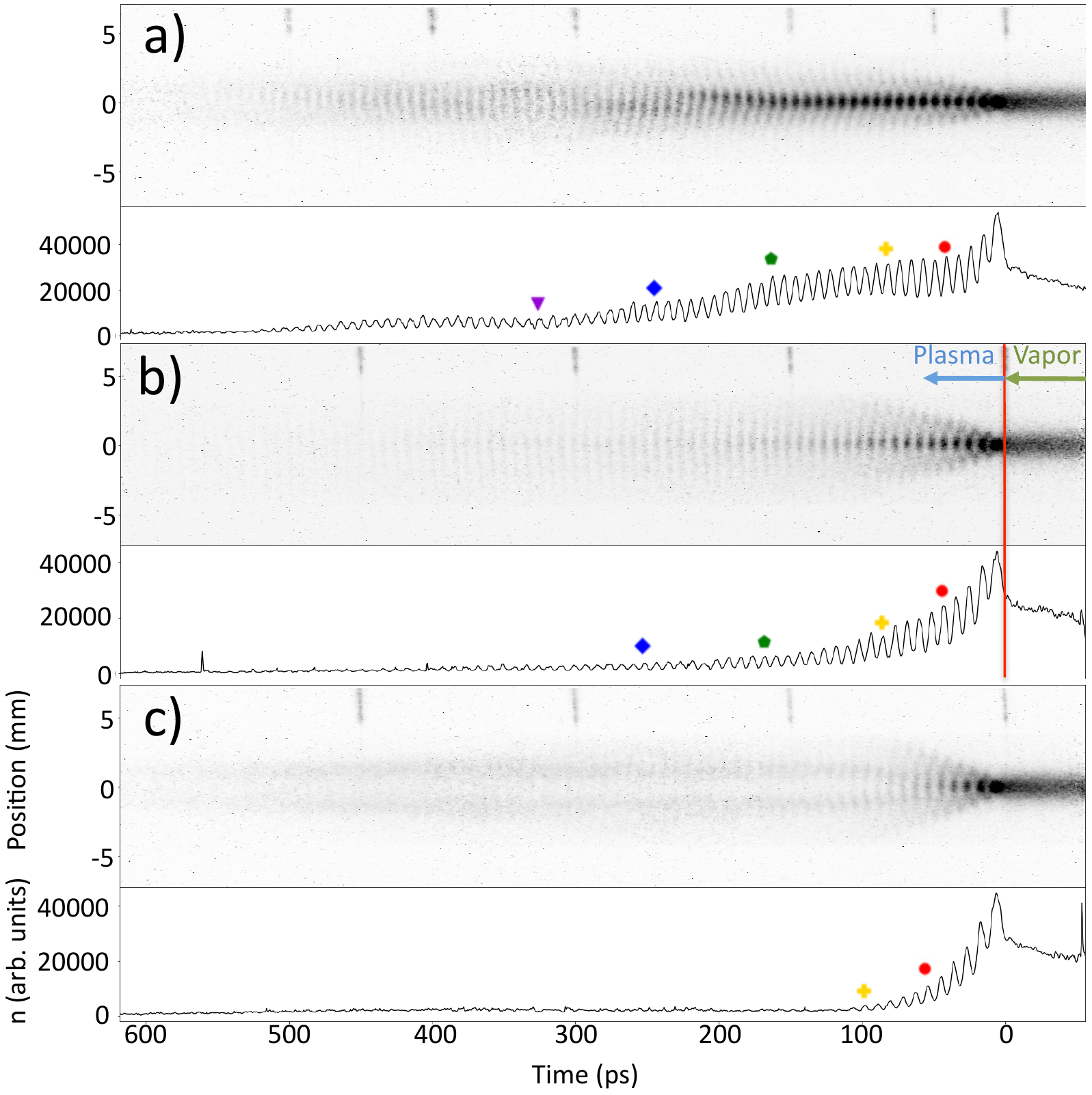}}
 \caption{\label{figStitchedStreak}Images of the proton bunch transverse density distribution (linear scale) as a function of time for three plasma density gradient values a) $g$=+1.99, b) +0.03 and c) -1.93\%/m. %
The bunch front is on the right hand side. %
Identical color table for all images. %
The ionizing laser pulse (not visible) is at $t$=0\,ps and its location marked by the red line on Fig. b). %
The part of the bunch at $t$$>$0\,ps propagated in the plasma and self-modulation is clearly visible. %
The vertical stripes in each image at position $>$4\,mm are the laser marker pulses used as time references~\cite{FabianStitchedMeth}. %
Time profiles of the density distribution (n) along the bunch axis (within $\pm\sigma_{r0}$ of the un-modulated bunch, black line) are shown below each image. %
Color symbols indicate micro-bunches number 5, 10, 20, 30, 40 used for Fig.~\ref{fig:ChargevsGrad} a). %
}
\end{figure} %

Figure~\ref{figStitchedStreak} shows time-resolved images of the bunch density with three of eight different plasma density gradient values for which data was acquired (see below). 
To obtain these $\cong$620\,ps-long images, several images with an acquisition time of 209\,ps and time resolution of $\sim$3\,ps were stitched together using a marker laser pulse as time reference, as described in~\cite{FabianStitchedMeth}. %
Marker laser pulses are visible at the top of Fig.~\ref{figStitchedStreak} images. %
To evidence modulation, we plot for each image a time profile of the bunch train, obtained by summing the counts on the image over a rms transverse range $\pm\sigma_{r0}$ about the bunch axis. 
Only the part of the bunch propagating in the plasma ($t$$>$0 with respect to the seed point) exhibits modulation. %
Images clearly show that with the positive density gradient ($g$=+1.99\%/m Fig.~a)) there are more micro-bunches along the train. %
This is confirmed by the corresponding time profiles. %
In the negative gradient value case (Fig.~c)) there are no visible micro-bunches at times later than $\sim$130\,ps ($\approx$0.6$\sigma_{t}$), while they are visible up to $\sim$250\,ps ($\approx$1.1$\sigma_{t}$) 
in the case of (essentially) constant plasma density (Fig.~b)), and up to $\sim$350\,ps ($\approx$1.5$\sigma_{t}$) with the positive gradient value. %
With the largest negative gradient value ($g$=-1.93\%/m), 
there is no charge visible along the bunch axis for $t$$>$130\,ps. %
The defocused protons are clearly visible away from the axis (Fig. c), $t$$>$150\,ps) with a modulation pattern with the wakefields' periodicity. %
Images with gradient values near $g$=0\%/m (e.g., Fig.~\ref{figStitchedStreak} b)) appear to have very symmetric and regular charge distribution. %
With the largest positive gradient ($g$=+1.99\%/m, Fig.~\ref{figStitchedStreak} a)) the charge density distribution exhibits significant irregularities. %
\begin{figure}[!ht]
\centerline{\includegraphics[width=0.8\linewidth]{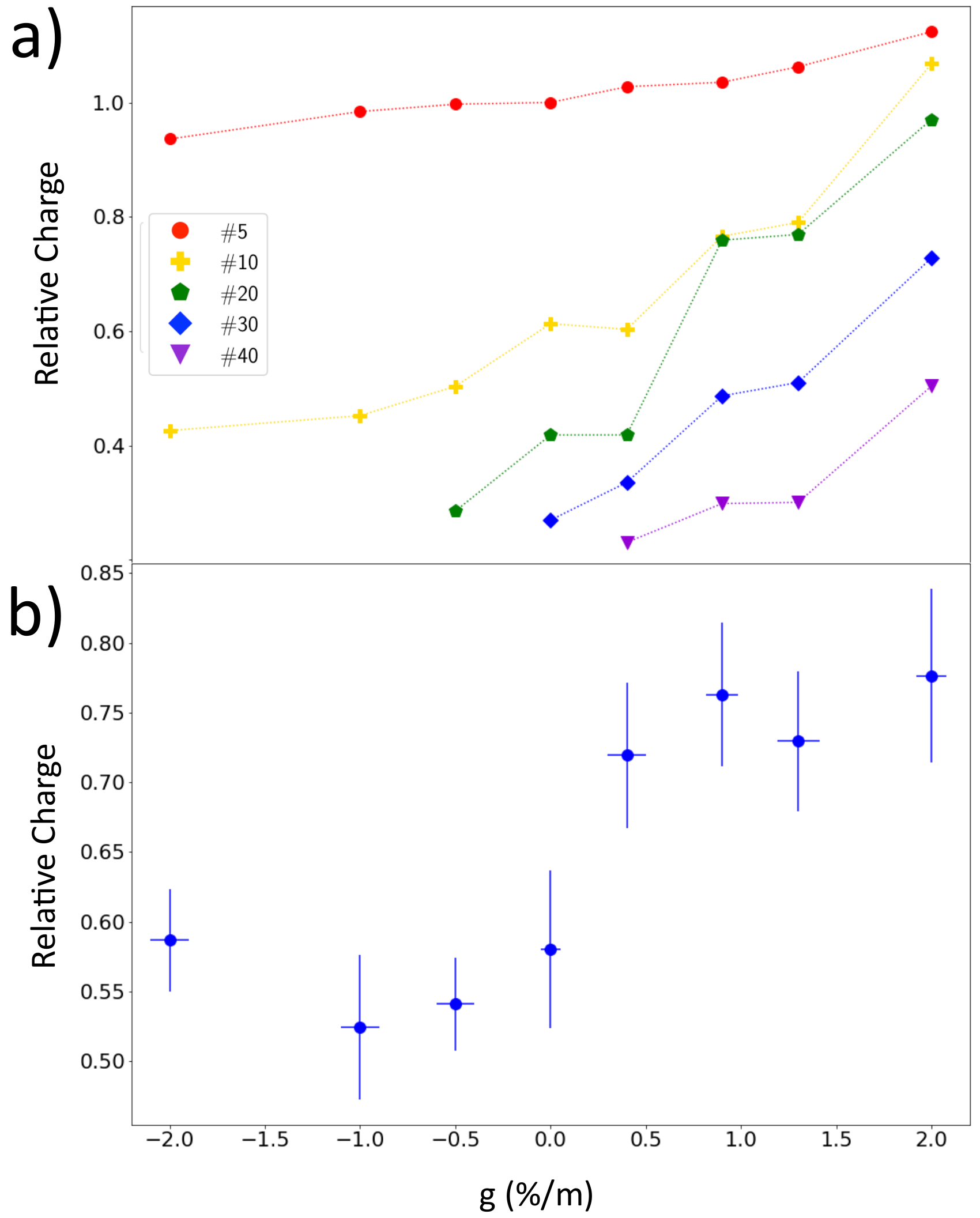}}
\caption{\label{fig:ChargevsGrad}a) Relative charge in micro-bunches number 5, 10, 20, 30, 40 (indicated by symbols of the same color as on Fig.~\ref{figStitchedStreak}, lines joining points to guide the eye) from the ionization front, versus plasma density gradient $g$. %
Charge in each micro-bunch is normalized to that in number 5 for $g$=+0.03\%/m density gradient. %
b) Relative charge in the bunch core (from time integrated images) versus plasma density gradient $g$. %
The charge is normalized to the average charge without plasma. %
Error bars correspond to standard deviations of the observations. %
}
\end{figure} %

In order to quantify the charge remaining along the self-modulated bunch, Fig.~\ref{fig:ChargevsGrad} a) shows the relative charge contained in micro-bunches number 5, 10, 20, 30 and 40 behind the seed point, as a function of $g$, and calculated according to the procedure detailed in Ref.~\cite{bib:annaeaac}. %
One can clearly observe that for all these micro-bunches, the charge is in/decreasing with in/decreasing gradient values. %
Further along the bunch, micro-bunches and charge are present on axis only with large positive gradient values, e.g., bunches number 30 and 40 and $g$$>$+1\%/m. %

A similar trend for the bunch relative charge along its axis as a function of gradient is obtained from time integrated images of the transverse charge distribution~\cite{MarleneTwoScreen17}. 
Figure~\ref{fig:ChargevsGrad} b) shows the bunch relative charge measured in one rms radius around the bunch center, i.e., its core. 
The charge contained in the core consists of the charge ahead of the seed point ($\cong$29\%), therefore unaffected by the SM process, and of that of the micro-bunches and thus remaining near the axis. 
The charge not in the core is found in the halo surrounding the core~\cite {bib:turner}. %
The charge in the core generally in/decreases as a function of in/decreasing gradient, except for large negative values ($g$$<$-1.0\%/m). %
This measurement is consistent with the results from time-resolved images, i.e., more charge in micro-bunches or core with more positive $g$ values. 

We note here that an estimate for the expected charge near the bunch is that a maximum of 50\% could remain after the seed point, assuming linear theory fields~\cite{bib:keinings} with half of the wakefields' period focusing for protons. %
With a Gaussian longitudinal profile and a seed point at time $t_{seed}$=+128\,ps along the bunch a fraction of $\cong$64\% would remain in the core. %
An optimum modulation would be with charge remaining only in the decelerating and focusing phase of the wakefields. %
That is only a quarter period in linear plasma wakefields' theory and thus a fraction of 47\% would remain. %
Figure~\ref{fig:ChargevsGrad} b) shows that without gradient the charge fraction remaining is $\sim$57\%. %
Measured values ($\cong$52 to $\cong$77\%) are a bit larger than expected ones, possibly because the ideal bunching is not reached and because charge from the defocused regions remains within the rms radius of the bunch. %

We determine the bunch modulation frequency $f_{CTR}$ with the CTR diagnostic~\cite{CTRdiagEAACNiMa18}. %
The CTR spectrum of a periodic bunch train consists of peaks at the modulation frequency and at its harmonics. %
We use heterodyne systems operating near the fundamental ($\sim$120\,GHz) and the second harmonic ($\sim$240\,GHz) frequencies of the modulation. %
Bench measurements show that these systems measure frequencies with $<$1\,GHz accuracy~\cite{CTRdiagEAACNiMa18}. %
The presence of harmonics in the signal indicates deep modulation of the charge density along the bunch, an indication of the nonlinear stage reached by the SM process, as can be seen in the profiles of Fig.~\ref{figStitchedStreak}. %
Time profiles are shallower than in reality due to the limited streak camera time resolution ($\sim$3\,ps). %

The discrete Fourier transform (DFT) frequency analysis used in~\cite{bib:karl} is also applied to the time-resolved images (Fig.~\ref{figStitchedStreak} and others) to obtain the modulation frequency $f_{Streak}$~\cite{bib:DFT}. %

\begin{figure}[!ht]
\centering     
\centerline{\includegraphics[width=0.7\linewidth]{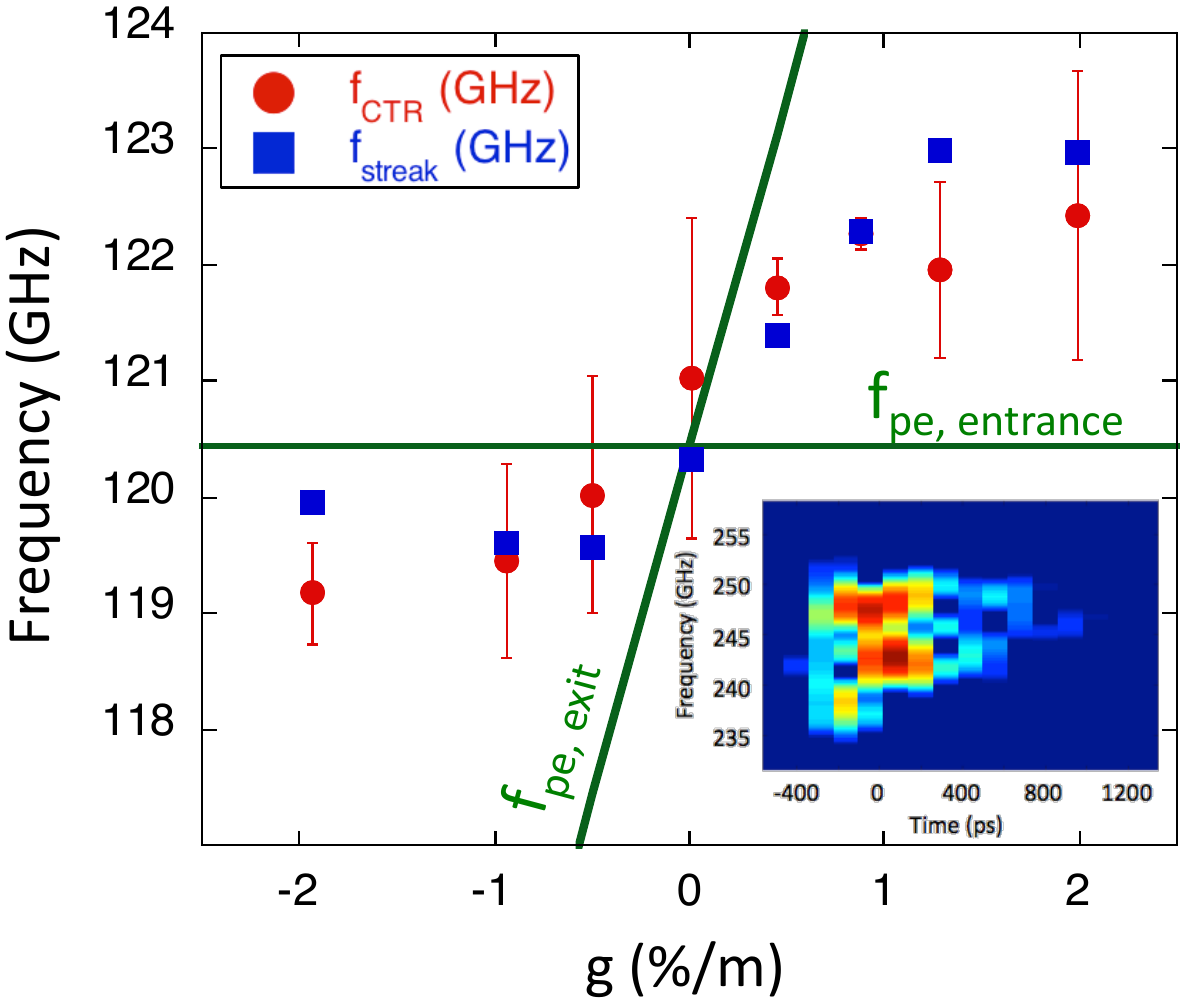}}
\caption{Proton bunch modulation frequency from CTR signals ($f_{CTR}$, red symbols, fundamental) and discrete Fourier transform of the streak camera profiles (as on Fig.~\ref{figStitchedStreak}) ($f_{Streak}$ blue symbols) versus plasma density gradient $g$. %
Error bars on $f_{CTR}$ indicate the standard deviation of the measurements at each gradient. %
The precision of $f_{Streak}$ is the bin size for the DFT, $\pm$0.27\,GHz. %
Green lines indicate the plasma frequencies at the plasma entrance ($f_{pe, entrance}$, constant) and at the exit ($f_{pe, exit}$) calculated from the plasma densities at these two locations. %
Inset: Spectrogram (frequency versus time~\cite{spectrogram}) of the CTR signal recorded at the second harmonic for an event with appearance of two distinct frequency peaks obtained with $g$=+1.98\%/m. %
Signal strength proportional to color, from blue to red. %
Detected frequencies are 242.8 and 247.7\,GHz.
}\label{fig2CTRfreqs}
\end{figure}

Figure~\ref{fig2CTRfreqs} shows these two sets of measured modulation frequencies as a function of $g$. 
There is very good agreement between the two independent measurement results. %
The modulation frequencies generally in/decrease with in/decreasing $g$ values and saturate at the largest absolute values. %
Measured frequencies are between the plasma frequencies determined from the entrance and exit plasma densities ($f$=$f_{pe}$=2$\pi/\omega_{pe}$$\propto$n$_{e0}^{1/2}$). %
With the largest $g$ values, the mean measured frequencies ($f_{CTR}$=119.2\,GHz, $f_{Streak}$=120.0\,GHz for $g$=-1.93\%/m, $f_{CTR}$=122.4\,GHz, $f_{Streak}$=123.0\,GHz for $g$=+1.99\%/m) are significantly different from plasma frequencies at the end of the plasma, 
108.2 and 131.9\,GHz, respectively. %

With large positive or negative gradient values ($\left|g\right|$$>$1\%/m), 
we observe two distinct frequency peaks in the CTR power spectrum, as shown for a single event in the spectrogram inset of Fig.~\ref{fig2CTRfreqs}~\cite{spectrogram}. %
This frequency measurement was performed at the second harmonic of f$_{CTR}$. 
The corresponding frequencies at the fundamental are $\cong$121.4 and $\cong$123.8\,GHz. %
The density gradient constantly detunes the resonance between the bunch modulation frequency and the local plasma frequency. %
This detuning generates an amplitude beating that has been observed in simulation results with plasma density gradients~\cite{bib:lotovbeat}. %
For excitation over a finite amount of time, this beating of the two frequencies becomes detectable when their separation is at least on the order of the inverse of the beating duration, here: $\Delta f\sim$1/($\sigma_{t}+t_{seed}$)=2.8\,GHz. %
In this case $\Delta f\cong$2.4\,GHz satisfies this condition. %


A higher/lower modulation frequency at the end of the plasma than at the entrance for positive/negative density gradient is direct evidence of the expected effect: the change in wakefields' phase velocity. %
Indeed, a higher/lower frequency corresponds to wakefields contracting/stretching along the bunch and plasma, with respect to the fixed seed point. %
This effect adds to the stretching away from the seed point occurring in a uniform density plasma where the wakefields' phase velocity is slower than the protons' velocity. 
A larger positive/negative phase velocity of wakefields along the bunch (in the bunch frame) means that protons spend less/more time in the passing defocusing fields and fewer/more of them are thus lost from the bunch train. %
This effect is stronger further along the bunch and explains the lengthening/shortening of the bunch train and gain/loss of charge observed with positive/negative $g$ values. %

Theoretical calculations show that, during the growth process, the wakefields' phase velocity at position $\xi=z-ct$ 
along the bunch and $z=ct$ along the plasma with electron density $n_{e0}$ differs from the particles or bunch velocity $v_b$ by~\cite{bib:phasevelocity}:
\begin{equation}
\Delta v_{\xi}\cong\frac{1}{2}\left(\frac{\xi}{ct}\right)^{1/3}\left(\frac{n_bm_e}{n_{e0}m_p\gamma_p}\right)^{1/3}v_b.
\label{eqn:vpukhov}
\end{equation}
The wakefields' phase velocity is slower than that of the bunch during the growth of the SM process. %
This difference increases along the bunch. %

In a plasma with a density gradient, a point of constant phase $\phi(\xi)=\frac{\omega_{pe}}{c}\xi$, initially at position $\xi$ along the bunch behind the seed point, moves closer to ($g>$0), or further ($g<$0) from the seed point. %
In the case of a linear plasma density gradient this corresponds to a velocity:
\begin{equation}
\Delta v_{\xi}=-\frac{1}{2}\xi\frac{(g/100)}{(1+(g/100)z)^{1/2}}v_b
\label{eqn:ppgrad}
\end{equation}
Possible compensation of the two opposite phase velocities with $g$$>$0 strictly occurs only for a given plasma length (at $z$) and position along the bunch (at $\xi$). %
Also, the first effect (Eq.~\ref{eqn:vpukhov}) only occurs over the growth length of the SM process, while the second one occurs all along the plasma. 
The overall effect is still to decrease the difference between the bunch and wakefields' velocities all along the bunch, as shown in Fig.~4 of Ref.~\cite{bib:phasevelocity}. %
One can thus expect it to lead to less overall dephasing, to more charge and micro-bunches to be preserved and possibly, for some parameters, to larger amplitude wakefields to be driven. %
However, this can occur only over a distance on the order of the growth length of the SM process. %
At longer distances the positive velocity of Eq.~\ref{eqn:ppgrad} has the same deleterious effect on the SM process as that that a positive gradient can counter. %
The effect of the negative gradient and negative wakefields' velocities has a compounding deleterious effect on the bunch train, as seen in the experimental data. %
The asymmetry of the effect of the plasma density gradient confirms that the phase velocity of the wakefields is slower than that of the protons~\cite{bib:phasevelocity,bib:schroeder}. %

Globally, the effect of a density gradient is to suppress the SM process growth~\cite{bib:schroedergrad}. %
According to Ref.~\cite{bib:schroedergrad} and within the assumptions of the model, the reduction in final relative radial modulation (assumed to be small) would be of about a factor three with respect to the constant density case, with a plasma density of 10$^{15}$\,cm$^{-3}$ and $g$=+2\%/m. %
This suppression occurs because of relative dephasing between the wakefields and the micro-bunching, as explained above. %
In order to determine the effect of the number of micro-bunches and of their charge on wakefields' amplitude, one must determine the longitudinal and transverse distribution of the protons in the wakefields. %
This is not possible without simultaneous measurement of the electron plasma density perturbation driven by the train. %
However, Fig.~2 of Ref.~\cite{bib:schroedergrad} does show a slight increase in SM growth for small gradient values, similar to those for which energy gain increase was observed in~\cite{bib:AWAKEelAccNature,bib:AWAKERoyal}. %
This increase in growth could result in shorter saturation length and longer acceleration length and/or larger wakefields' amplitude. %
In~\cite{bib:AWAKEelAccNature,bib:AWAKERoyal} the density gradient may also have affected the electrons' capture process. %

We presented experimental results on the effect of linear plasma density gradients on the self-modulation of a 400\,GeV proton bunch. %
Results show that, when compared to the no or negative gradients cases, positive density gradients up to 2\%/m along the beam path increase the number of micro-bunches observed, 
as well as their charge. 
Density gradients also lead to modulation frequency shifts 
with saturation at large values (positive or negative). %
Two frequencies appear in the frequency spectra at high gradients ($g\sim\pm$2\%\,m). 
These observations are consistent with modification of the wakefields' phase velocity, leading to fewer/more protons being defocused for $g$$>$0\%/m or $g$$<$0\%/m. %
Energy gain measurements~\cite{bib:AWAKEelAccNature} suggest that modest gradients value (up to $g$=+0.2\%/m) lead to larger energy gain than with a plasma with constant density ($g$=0\%/m). %
Capture of electrons at larger gradient values decreased with this injection scheme, thus preventing observation of possible decrease in energy gain. %

It is clear that a density gradient cannot be sustained over very long plasma lengths and could be beneficial only over the saturation length of the self-modulation process~\cite{bib:turnersat} (as far as this process and not the acceleration one is concerned). %
A positive density step, sharp~\cite{bib:lotovstep} or continuous~\cite{bib:minakov}, has so far been chosen for the Run 2 of AWAKE to maintain wakefields' amplitudes at large values over longer distances for larger energy gain than in a constant density plasma~\cite{bib:muggliRun2}. %

\section*{Acknowledgements}
This work was supported in parts by a Leverhulme Trust Research Project Grant RPG-2017-143 and by STFC (AWAKE-UK, Cockroft Institute core and UCL consolidated grants), United Kingdom; a Deutsche Forschungsgemeinschaft project grant PU 213-6/1 ``Three-dimensional quasi-static simulations of beam self-modulation for plasma wakefield acceleration''; the National Research Foundation of Korea (Nos.\ NRF-2016R1A5A1013277 and NRF-2019R1F1A1062377); the Portuguese FCT---Foundation for Science and Technology, through grants CERN/FIS-TEC/0032/2017, PTDC-FIS-PLA-2940-2014, UID/FIS/50010/2013 and SFRH/IF/01635/2015; NSERC and CNRC for TRIUMF's contribution; the U.S.\ National Science Foundation under grant PHY-1903316; the Wolfgang Gentner Programme of the German Federal Ministry of Education and Research (grant no.\ 05E15CHA); and the Research Council of Norway. M. Wing acknowledges the support of the Alexander von Humboldt Stiftung and DESY, Hamburg. Support of the Wigner Datacenter Cloud facility through the ”Awakelaser” project and the support of P\'{e}ter L\'{e}vai is acknowledged. The work of V. Hafych has been supported by the European Union's Framework Programme for Research and Innovation Horizon 2020 (2014--2020) under the Marie Sklodowska-Curie Grant Agreement No.\ 765710.  The AWAKE collaboration acknowledge the SPS team for their excellent proton delivery.



\begin{thebibliography}{99}
\bibitem{bib:cst}Constants have usual meaning: $e$   electron charge, $\epsilon_0$  vacuum permittivity, $m_e$  electron mass, $c$  speed of light in vacuum.
\bibitem{bib:DFT} We note here that the entire image width  was used for the DFT. Analysis of narrower image widths yields more complete results on the modulation frequency. These will be detailed in an upcoming manuscript.
\bibitem{spectrogram}Frequency of the first harmonic of CTR obtained from discrete Fourier transform of a heterodyne system signal f$_{IF}$=$\left|f_{CTR}-n\cdot f_{LO}\right|$, with harmonic number n=12, in 130\,ps long successive time windows.
\end{thebibliography}
%

\end{document}